\documentclass[twocolumn,showpacs,preprintnumbers,amsmath,amssymb,prd,nofootinbib]{revtex4-2}
\usepackage[utf8]{inputenc} 
\usepackage{amsmath}
\usepackage{amssymb}
\usepackage{bm}
\usepackage{graphicx}
\usepackage[justification=raggedright,singlelinecheck=false]{caption}
\usepackage{subcaption}
\usepackage{mathrsfs} 
\usepackage[mathscr]{euscript}
\usepackage{multirow}
\usepackage[normalem]{ulem}
\usepackage{graphicx}
\usepackage{booktabs}
\usepackage{xcolor, color, framed}
\usepackage[colorlinks, linkcolor=red,anchorcolor=green,citecolor=blue]{hyperref}
\usepackage{slashed}
\usepackage{ulem}

\graphicspath{{./confirmed_figure/}}

\usepackage{tikz,xcolor,hyperref}
\usepackage{ctable}

\definecolor{lime}{HTML}{A6CE39}
\DeclareRobustCommand{\orcidicon}{
	\begin{tikzpicture}
	\draw[lime, fill=lime] (0,0) 
	circle [radius=0.16] 
	node[white] {{\fontfamily{qag}\selectfont \tiny ID}};
	\draw[white, fill=white] (-0.0625,0.095) 
	circle [radius=0.007];
	\end{tikzpicture}
	\hspace{-2mm}
}

\foreach \x in {A, ..., Z}{%
	\expandafter\xdef\csname orcid\x\endcsname{\noexpand\href{https://orcid.org/\csname orcidauthor\x\endcsname}{\noexpand\orcidicon}}
}

\newcommand{\colorSchemeChosen}{$t/a$ = 16 (blue), 17 (red) and $18$ (green)}

\begin{document}

\title{Probing \texorpdfstring{nucleon-$\Omega_{\rm ccc}$}{nucleon-Omegaccc}
 interaction via lattice QCD at physical quark masses}

\author{Liang Zhang\orcidA{}}
\email{zhangliang@sinap.ac.cn}
\affiliation{Shanghai Institute of Applied Physics (SINAP), Shanghai, China}
\affiliation{Key Laboratory of Nuclear Physics and Ion-beam Application (MOE), Institute of Modern Physics, Fudan University, Shanghai 200433, China}
\affiliation{RIKEN Center for Interdisciplinary Theoretical and Mathematical Sciences  (iTHEMS), Wako, 
351-0198, Japan}
\affiliation{School of Nuclear Sciences and Technology, University of Chinese Academy of Sciences, Beijing 100049, China}

\author{Takumi Doi\orcidB{}}
\email{doi@ribf.riken.jp}
\affiliation{RIKEN Center for Interdisciplinary Theoretical and Mathematical Sciences  (iTHEMS), Wako, 
351-0198, Japan}
\author{Yan Lyu\orcidC{}}
\email{yan.lyu@riken.jp}
\affiliation{RIKEN Center for Interdisciplinary Theoretical and Mathematical Sciences  (iTHEMS), Wako, 
351-0198, Japan}

\author{Tetsuo Hatsuda\orcidD{}}
\email{thatsuda@riken.jp}
\affiliation{RIKEN Center for Interdisciplinary Theoretical and Mathematical Sciences  (iTHEMS), Wako, 351-0198, Japan}
\affiliation{Kavli Institute for the Physics and Mathematics of the Universe (Kavli IPMU), WPI,  
The University of Tokyo, Kashiwa, 
277-8568, Japan}

\collaboration{HAL QCD Collaboration}

\author{Yu-Gang Ma\orcidE{}}
\email{mayugang@fudan.edu.cn}
\affiliation{Key Laboratory of Nuclear Physics and Ion-beam Application (MOE), Institute of Modern Physics, Fudan University, Shanghai 200433, China}
\affiliation{Shanghai Research Center for Theoretical Nuclear Physics, NSFC and Fudan University, Shanghai 200438, China}

\begin{abstract}
We study the S-wave interactions between the nucleon ($N$) and the triply charmed Omega baryon ($\Omega_{\mathrm{ccc}}$) using (2+1)-flavor lattice QCD with a physical pion mass ($m_\pi \simeq 137.1$ MeV) on a lattice volume $\simeq (8.1~\mathrm{fm})^3$. 
The charm quark is implemented with a relativistic heavy-quark action at its physical mass. 
Employing the time-dependent HAL QCD method, the $N$-$\Omega_{\mathrm{ccc}}$ potentials in the spin-1 ($^3\mathrm{S}_1$) and spin-2 ($^5\mathrm{S}_2$) channels are extracted.
In both channels, overall attraction is found with the scattering parameters,
$a_0 = 0.56(0.13)\left(^{+0.26}_{-0.03}\right)$ fm and $r_{\mathrm{eff}} = 1.60(0.05)\left(^{+0.04}_{-0.12}\right)$ fm 
for the $^3\mathrm{S}_1$ channel, and 
$a_0 = 0.38(0.12)\left(^{+0.25}_{-0.00}\right)$ fm and $r_{\mathrm{eff}} = 2.04(0.10)\left(^{+0.03}_{-0.22}\right)$ fm
for the $^5\mathrm{S}_2$ channel,
indicating the absence of a dibaryon bound state.
The extracted potentials are further decomposed into spin-independent and spin-dependent components. 
The spin-independent potential is a dominant component and features a short-range attractive core and a long-range attractive tail, while the spin-dependent potential shows short-range attraction (repulsion) in the spin-1 (spin-2) channel.
Qualitative comparisons with previous studies of the $N$-$J/\psi$ and $N$-$\Omega_{\rm{sss}}$ systems at $m_\pi \simeq 146$ MeV are provided, emphasizing the role of heavy-hadron chromo-polarizability arising from soft-gluon exchange between the nucleon and flavor-singlet hadrons. 
The charm quark mass dependence of the $N$-$\Omega_{\rm ccc}$ potential is investigated as well. 
\end{abstract}

\maketitle

\section{Introduction}
\label{sec:intro}

{Quantum Chromodynamics (QCD) is the fundamental theory of strong interaction, governing the dynamics of quarks and gluons. In the low-energy regime, its strongly coupled, non-perturbative nature renders analytical calculations intractable. Lattice QCD (LQCD), leveraging advances in numerical algorithms and computational power, enables first-principles investigations not only of single hadrons but also of the hadronic interactions~\cite{Luscher:1990ux,Ishii:2006ec}.}
 
{Among various two-hadron systems, the di-hadrons with distinct quark flavor content have been investigated using the time-dependent HAL QCD method \cite{Ishii:2012ssm}: $N$-$\Omega_{\rm{sss}}$ \cite{HALQCD:2018qyu}, $N$-$\phi$ \cite{Lyu:2022imf}, $N$-$J/\psi$ and $N$-$\eta_{\rm c}$ \cite{Lyu:2024ttm}.  They are of particular interest, as they offer a unique opportunity to investigate long-range spin-isospin independent interactions as well as  short-range interactions in the absence of quark Pauli blocking. 
It has been found that all of these channels exhibit fully attractive interactions from short to long distances. 
Especially, the $N$-$\Omega_{\rm{sss}}$ system in the $^5{\rm S}_2$ channel is found to have a quasi-bound state near unitarity with the binding energy of about 2 MeV \cite{HALQCD:2018qyu}.
}

{Based on the lattice QCD results, the $N$-$\Omega_{\rm{sss}}$ system has also been intensively studied via femtoscopic correlation measurements in heavy-ion and proton-proton collisions~\cite{STAR:2018uho,ALICE:2020mfd}, which provide evidence for an attractive interaction.
Inspired by these results, a quark-model study suggested that the heavier $N$-$\Omega_{\rm ccc}$ system might form a bound state~\cite{Huang:2019esu}. 
To this end, first-principles calculations can provide a valuable theoretical prediction for such triply charmed states. }

{Unlike the $N$-$\Omega_{\rm sss}$ system, which lies above the $\Lambda$-$\Xi$ and $\Sigma$-$\Xi$ thresholds, the $N$-$\Omega_{\rm ccc}$ threshold (approximately 5740~MeV) sits well below both the $\Lambda_{\rm c}$-$\Xi_{\rm cc}$ threshold (5910~MeV) and the $\Sigma_{\rm c}$-$\Xi_{\rm cc}$ threshold (6080~MeV), making it the lowest state among all $C$(charm number)=3 dibaryon systems. Such a $\sim 200$~MeV energy gap provides an ideal, uncontaminated setting to examine the low‐energy $N$-$\Omega_{\rm ccc}$ interactions.}

{The primary purpose of this paper is to deliver the first investigation of the $N$-$\Omega_{\rm{ccc}}$ interaction in the two spin-channels ($^3{\rm{S}}_1$ and $^5{\rm{S}}_2$), employing $(2+1)$-flavor lattice QCD with physical light-quark masses ($m_{\pi}\simeq137.1\ \mathrm{MeV}$) and the physical charm-quark mass $\bigl(m_{\eta_{\rm{c}}}+3m_{J/\psi}\bigr)/4\simeq 3068.5\ \mathrm{MeV}$. We extract the $N$-$\Omega_{\rm{ccc}}$ interaction potentials using the HAL QCD method \cite{Ishii:2006ec,Ishii:2012ssm,Aoki:2020bew}, a non-perturbative framework that determines the interaction directly from  spacetime correlation functions of multiple hadrons.}

Based on lattice QCD data, one can determine the quark-mass dependence of the interaction between $N$-$\Omega_{\rm{ccc}}$ and $N$-$\Omega_{\rm sss}$, providing valuable insights into heavy-hadron interactions. Notably, since the $N$-$\Omega_{\rm{ccc}}$ lies below the $\Lambda_{\rm c}$-$\Xi_{\rm cc}$ and $\Sigma_{\rm c}$-$\Xi_{\rm cc}$ thresholds, it is possible to disentangle spin-dependent and spin-independent forces from lattice data in the ${}^3{\rm S}_1$ and ${}^5{\rm S}_2$ channels, an analysis that is not feasible in the case of $N$-$\Omega_{\rm sss}$~\cite{HALQCD:2018qyu}. Furthermore, a comparison between the long-distance behaviors of the $N$-$\Omega_{\rm ccc}$ and $N$-$J/\psi$ systems offers important information on the chromo-polarizability of heavy hadrons. 
In addition, we study the charm quark mass dependence of  the $N$-$\Omega_{\rm ccc}$ potential.

This paper is organized as follows. 
Sec.~\ref{sec:HALmethod} introduces the HAL QCD method for extracting the hadron interactions from lattice QCD. 
Sec.~\ref{sec:latticeset} summarizes the setup of our lattice QCD simulations at the physical point. 
A detailed analysis of the $N$-$\Omega_{\rm ccc}$ interactions in the $^3{\rm S}_1$ and $^5{\rm S}_2$ channels is presented in Sec.~\ref{sec:res}. 
Sec.~\ref{sec:disc} discusses the underlying mechanisms driving the $N$-$\Omega_{\rm ccc}$ interactions. 
Finally, Sec.~\ref{sec:summary} provides a summary and concluding remarks.

\section{HAL QCD method}
\label{sec:HALmethod}
Let us start with the following  correlation function for $N$ and $\Omega_{\rm ccc}$ (or $\Omega_{\rm 3c}$  in short):
\begin{align}
\label{eq:4pt-corr}
    F^J\left(\bm{r},t\right) &=  
\sum_{\bm{x}}\left\langle0\right|\left[N(\bm{x},t){\Omega_{\rm 3c}}(\bm{r}+\bm{x},t)\right]^J \bar{\mathcal{J}}^{J}_{N\Omega_{\rm 3c}}(0)\left|0\right\rangle \notag\\
 &=\sum_{n} A^J_n\psi_{N\Omega_{\rm 3c}}^{J,n}(\bm{r})e^{-W_n t},
\end{align}
where $\bar{\mathcal{J}}^{J}_{N\Omega_{\rm 3c}}(0)$ is an operator at time 0 (the source), projected onto a total-spin $J$ state, constructed from a wall-type six-quark source operator with zero angular momentum ($L=0$).
The nucleon and $\Omega_{\rm 3c}$ operators at time $t$ (the sink)  are defined as,
\begin{equation}
\label{eq:Boperator}
\begin{aligned}
   {}[N]_{\alpha}(\bm{x}) &=\varepsilon_{abc}\left({u^a}^{T}(\bm{x}){\rm C}\gamma_5 d^b(\bm{x})\right)q^c_\alpha(\bm{x}),
\\
   [{\Omega_{\rm 3c}]_{\beta,{l}}(\bm{x})}&=\varepsilon_{abc}\left({Q^a}^T(\bm{x}){\rm {C}}\gamma_{l}Q^b(\bm{x})\right)Q^{c}_{\beta}(\bm{x}) ,
\end{aligned}    
\end{equation}
where $q= (u,d)^T$ denoting the light quark field, $Q$ being the heavy charm quark field, $\alpha$ and $\beta$ are spinor indices, $l$ is a spatial index for the gamma matrix, and $a,b,c$ denote color indices. 
${\rm C} = \gamma_4 \gamma_2$ is the charge conjugation matrix.
They are combined into the operator with $(J, s, L) = (s,s,0)$ by 
the spacial projection $P_{A_1^+}$ onto the $A_1^+$ irreducible representation of the cubic group
and $P^{s}_{\alpha\beta,{l}}$ onto the spin $s$ state. 
The explicit form of the spin projection operator reads  
\begin{equation}
    \begin{aligned}
        P^s_{\alpha\beta,l}=\sum_{m_l}&\langle\frac{1}{2},m_{\beta};1,m_l|\frac{3}{2},m_{\beta}+m_l\rangle\times\\
        &\langle\frac{1}{2},m_{\alpha};\frac{3}{2},m_{\beta}+m_l|s,m_{\alpha}+m_{\beta}+m_l\rangle M_{m_{l}l},
    \end{aligned}
\end{equation}
where  $\alpha,\beta \in \{1,2\}$ and $l\in\{1,2,3\}$ denote  the spinor and the spatial indices, respectively,
 with $m_{\alpha,\beta} \in \{+\frac{1}{2}, -\frac{1}{2}\}$ and $m_l \in \{+1, 0, -1\}$.
Also, the Clebsch-Gordan coefficients are denoted by $\langle s_1,m_1;s_2,m_2|s,m\rangle$ and the transformation matrix $M$ converts spatial indices to spherical basis components:
\begin{equation}
    M =\frac{1}{\sqrt{2}} \begin{pmatrix}
        -1 &i & 0 \\
        0                   & 0                   & \sqrt{2} \\
        1  & i & 0
    \end{pmatrix}.
\end{equation}

Denoting  $\left|W_n\right\rangle$ as the $n$-th energy eigenstate with energy $W_n$ of the 6-quark system containing three charm quarks, 
$A^J_n\equiv\left\langle W_n\right|
 \bar{\mathcal{J}}^{J}_{N\Omega_{\rm 3c}}(0)\left|0\right\rangle$, while $\psi_{N\Omega_{\rm 3c}}^{J,n}(\bm{r})\equiv P_{A_1^+}P^{s}_{\alpha\beta,{l}}\sum_{\bm{x}}\left\langle0\right|[N]_{\alpha}(\bm{x})[\Omega_{\rm 3c}]_{\beta,{l}}(\bm{r}+\bm{x})
 \left|W_n\right\rangle$ 
 is the S-wave Nambu-Bethe-Salpeter (NBS) wave function with spin $s$. 
 
For the purpose to 
extract the $N$-$\Omega_{\rm 3c}$ potential, we introduce the
``R-correlator'';
$    R^J(\bm{r},t)\equiv {F^J(\bm{r},t)}/\left[G_N(t)G_{\Omega_{\rm 3c}}(t)\right]$
with $G_B(t) \propto \exp(-m_B t)$ being  the single-baryon correlator.
$R^J(\bm{r},t)$ satisfies the following integrodifferential equation at low energies according to the time-dependent HAL QCD method  \cite{Ishii:2012ssm,Miyamoto:2017tjs}
\begin{widetext}
\begin{equation}
\label{eq:S-eq}
    \left(-\partial_t +\frac{1+3\delta^2}{8\mu}\partial_t^2+\mathcal{O}(\partial_t^3)-\hat{H}_0\right)R^J(\bm{r},t)\\
    =\int{d^3\bm{r}'U(\bm{r},\bm{r}')R^J}(\bm{r}',t),
\end{equation}
\end{widetext}
with the kinetic operator $\hat{H}_0\equiv-\nabla^2/2\mu $, the reduced mass $\mu\equiv m_Nm_{\Omega_{\rm 3c}}/(m_N+m_{\Omega_{\rm 3c}})$ and the asymmetry parameter $\delta\equiv (m_{\Omega_{\rm 3c}}-m_N)/(m_{\Omega_{\rm 3c}}+m_N)$. 
The higher-order temporal derivatives are negligibly small in our calculation.
The non-local potential $U(\bm{r},\bm{r}')$ is energy-independent \cite{Aoki:2012tk, Aoki:2009ji, Ishii:2006ec} and admits a derivative expansion: $U(\bm{r},\bm{r}')=V(\bm{r},\nabla)\delta^{(3)}(\bm{r}-\bm{r}')$. At the leading order (LO) of the expansion in terms of $\nabla$, the effective central potential with the tensor force absorbed into $V_0$ \cite{Aoki:2009ji} has the form
\begin{equation}
\label{eq:local-pot}
\begin{aligned}
    V^{J}_{\rm LO}(r)&=V_0(r)+\vec{s}_{_N}\cdot\vec{s}_{_{\Omega_{\rm 3c}}}V_{s}(r)\\
    &\simeq R^J(\bm{r},t)^{-1}\left(-\partial_t +\frac{1+3\delta^2}{8\mu}\partial_t^2-\hat{H}_0\right)R^J(\bm{r},t),
\end{aligned}
\end{equation}
Here $V_0$ and $V_s$ denote the spin-independent and spin-dependent potentials, respectively. 
The spin operators for the spin-1/2 nucleon and the spin-3/2 $\Omega_{\rm 3c}$ are denoted by $\vec{s}_{_N}$ (2$\times$2 matrices) and $\vec{s}_{_{\Omega_{\rm 3c}}}$  (4$\times$4 matrices), respectively. Spatial and temporal derivatives on each lattice site $(\bm{r}, t)$  are computed by a central difference scheme with nearest-neighbor points. The obtained potential enables the calculation of physical observables in the infinite volume, since it is well-localized in space.

To stay away from the inelastic two spin-$1/2$ baryons channels ($\Lambda_{\rm c}$-$\Xi_{\rm cc}$ and $\Sigma_{\rm c}$-$\Xi_{\rm cc}$), $t$ must be chosen sufficiently large. 
In this study, we select $t/a = 16-18$, a range ensuring that such inelastic effects become negligibly small.
Then we could successfully extract the potential not only for the spin-2 channel but also for the spin-1 channel, unlike the $N$-$\Omega_{\rm 3s}$ case where only the potential in spin-2 channel was accessible \cite{HALQCD:2018qyu}.

\section{Lattice setup}
\label{sec:latticeset}
We employ the ``HAL-conf-2023" gauge configuration set (``F-conf" in short) \cite{Aoyama:2024cko}, generated in $(2+1)$-flavor lattice QCD,  by using the supercomputer Fugaku at RIKEN. The simulations use the Iwasaki gauge action at $\beta = 1.82$ and the nonperturbatively $\mathcal{O}(a)$-improved Wilson quark action with stout smearing. Gauge configurations are generated on a $96^4$ lattice at physical quark masses, corresponding to a pion mass of $m_{\pi} \simeq 137.1$ MeV.
The lattice spacing is $a \simeq 0.084372$ fm ($a^{-1} \simeq 2338.8$ MeV), 
yielding a sufficiently large spatial extent of $L \simeq 8.1$ fm to accommodate two-baryon systems.

For charm quarks, we adopt the relativistic heavy quark (RHQ) action \cite{Aoki:2001ra} to eliminate leading- and next-to-leading-order discretization errors. Two RHQ parameter sets (Set1 and Set2) from Ref.~\cite{Namekawa:2017nfw} are employed to enable interpolation to the physical charm quark mass. These parameters reproduce the dispersion relation of the spin-averaged 1S charmonium state, defined as the mass-weighted average of the spin-singlet $\eta_c$ and spin-triplet $J/\psi$.

For the source operator $\mathcal{J}^{J}_{N\Omega_{\rm 3c}}(0)$, we employ a wall-type quark source with Coulomb gauge fixing. 
Quarks satisfy the periodic boundary condition in both of spatial and temporal directions.
Our analysis uses 1,600 gauge configurations, sampled at intervals of 5 trajectories across 5 independent runs. Statistical precision is further enhanced per configuration by averaging forward and backward propagations, applying hypercubic symmetry via 4 lattice rotations, and sampling 32 distinct source locations. This procedure generates 409,600 measurements per parameter set. Statistical errors are evaluated via the jackknife method with an 80-configuration bin size.
Comparison with 40-configuration bin size confirms that our results are insensitive to the bin size. 

To suppress residual contributions from higher angular momentum states ($L \geq 4$) after $A^+_1$ projection, we implement the Misner’s method \cite{Misner:1999ab} for approximate partial wave decomposition on the cubic lattice. This technique carries out the S-wave projection efficiently as described in Ref.~\cite{Miyamoto:2019jjc}.

{Single-hadron masses are extracted by a single-state fit to the correlation functions, with the fit ranges chosen from the plateau regions of the effective mass plots.
Table~\ref{tab:intwepolateTwoSet} lists the masses of the spin-averaged 1S charmonium state $\left(m_{\eta_{\rm{c}}}+3m_{J/\psi}\right)/4$ and the $\Omega_{\rm 3c}$ baryon 
($m_{\Omega_{\rm 3c}}$), calculated using Set1 and Set2 with single-state 
fits over the interval $t/a = 30-40$~\cite{Lyu:2025ncq}.}
Also included are values obtained from a linear interpolation $\left(0.2631 \times \text{Set1} + 0.7369 \times \text{Set2}\right)$ to match experimental data.
Our result for $m_{\Omega_{\rm 3c}}$ agrees numerically with prior determinations
obtained from $(2+1)$-flavor configurations by PACS-CS Collaboration \cite{PACS-CS:2013vie},  
the value calculated in di-$\Omega_{\rm 3c}$ study by the HAL QCD Collaboration \cite{Lyu:2021qsh}, which utilized ``K-conf" with $m_{\pi}\simeq 146.4$ MeV generated by the PACS Collaboration on the supercomputer K at RIKEN \cite{Ishikawa:2015rho}. {It is also consistent with the recent update from the F-conf ensembles~\cite{Lyu:2025ncq}.}
For the nucleon mass, we obtain $m_N=938.9(2.8)$ MeV from the interval $t/a = 14-20$~\cite{Aoyama:2024cko}, consistent with the lattice result from Ref.~\cite{Aoyama:2024cko} and the experimental value from PDG \cite{ParticleDataGroup:2024cfk}. 

\begin{table}
    \centering
    \caption{First and second rows show the spin-averaged mass of 1S charmonium $\left( \left(m_{\eta_{\rm{c}}}+3m_{J/\psi}\right)/4\right)$ 
    as well as the mass of $\Omega_{\rm 3c}$ ($m_{\Omega_{\rm 3c}})$ computed using Set1 and Set2 with  statistical errors.
    The third row presents interpolated values derived from Set1 and Set2, while the last row displays the experimental value.}
    \begin{tabular}{ccc} 
    \toprule
         &   $\left(m_{\eta_{c}}+3m_{J/\psi}\right)/4$ [MeV]& $m_{\Omega_{\rm 3c}}$ [MeV]\\ 
    \midrule
         Set1&  3101.9(0.1)& 4846.4(0.1)\\  
         Set2&  3056.6(0.1)& 4779.0(0.1)\\
         Interpolation&  3068.5(0.1)& 4796.8(0.1)\\  
         Experiment&  3068.5& {\rm none}\\
    \bottomrule
    \end{tabular}
    
    \label{tab:intwepolateTwoSet}
\end{table}

\section{Numerical results}
\label{sec:res}

\subsection{\texorpdfstring{$N$-$\Omega_{\rm 3c}$}{NOmegaccc} potential in different spin channels}
\label{sec:NOpot}
\begin{figure}[t]
\begin{subfigure}{0.99\linewidth} 
    \centering{
            \includegraphics[width=0.99\linewidth]{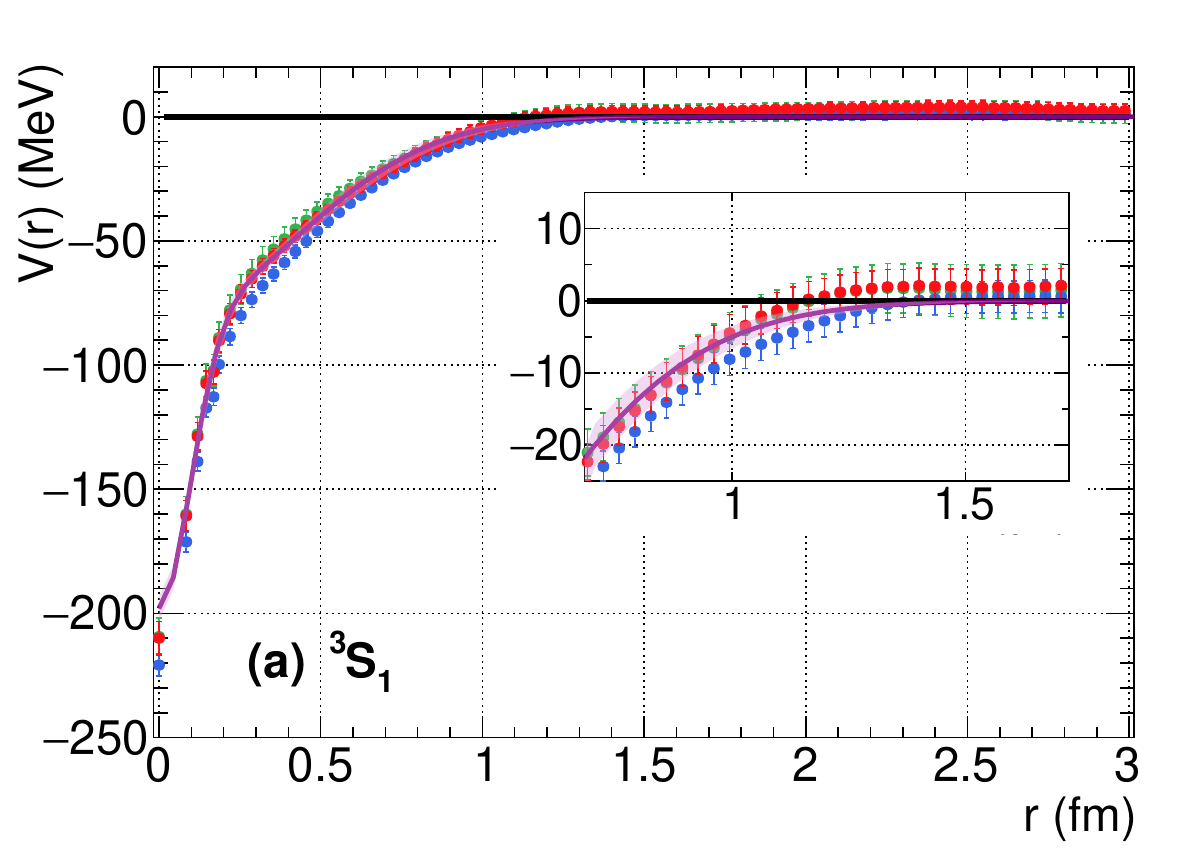}
    }

    \end{subfigure}
    \begin{subfigure}{0.99\linewidth} 
    \centering{
            \includegraphics[width=0.99\linewidth]{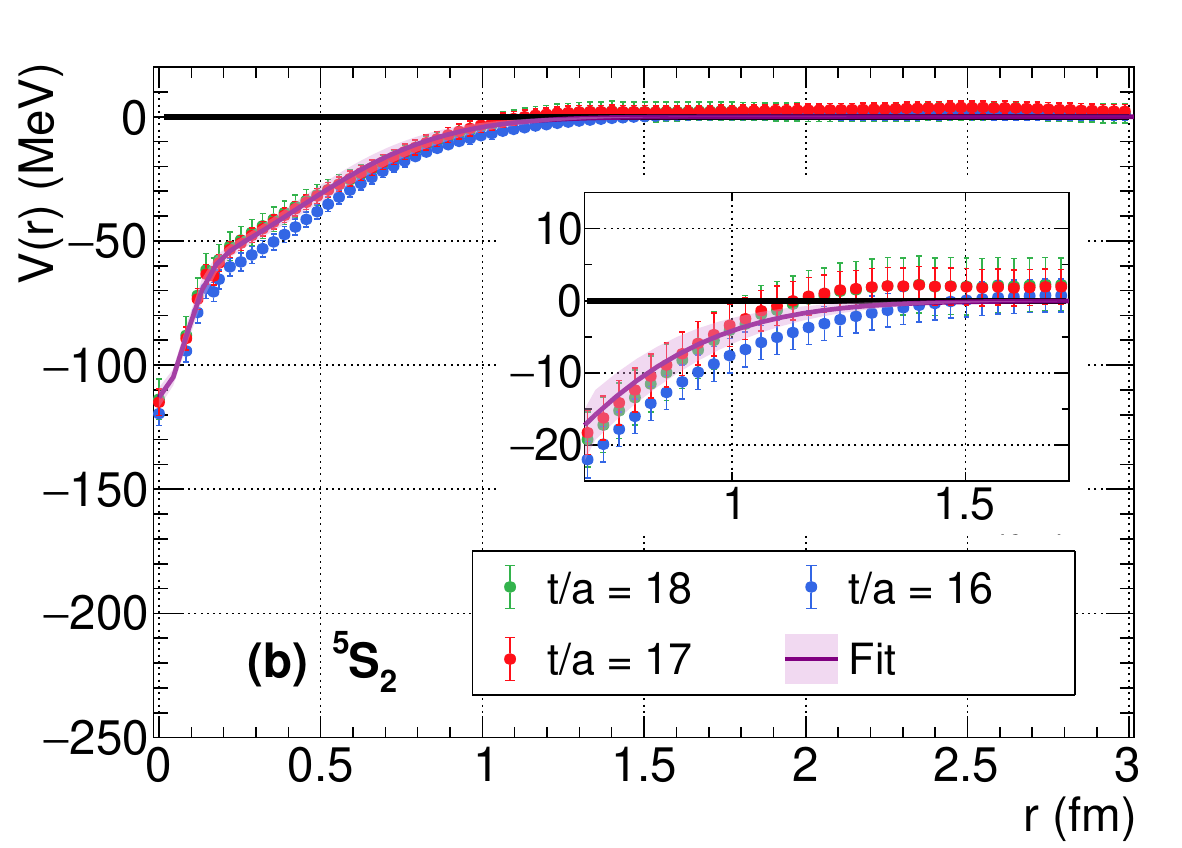}
    }

    \end{subfigure}
    \caption{ The $N$-$\Omega_{\rm 3c}$ potential in the $S$-wave extracted from lattice data at \colorSchemeChosen \ with statistical errors. The panel (a)  (the panel (b)) shows the potential in the $^3{\rm S}_1$ ($^5{\rm S}_2$) channel. 
    Fits with the two-range Gaussian $V_{\rm fit}(r)$ at $t/a=17$ in the range of $0<r<3$ fm  are drawn by the purple.
    }
    \label{fig:potential}
\end{figure}

The $N$-$\Omega_{\rm 3c}$ potentials in the ${}^3{\rm S}_1$ and ${}^5{\rm S}_2$ channels are presented in Fig.~\ref{fig:potential} for temporal separations $t/a = 16$, $17$, and $18$.
The results are obtained by interpolating the potentials from Set1 and Set2 to the physical charm quark mass, where the two sets of potentials are found to be almost identical, with their ratio exhibiting $1/m_{_{\Omega_{\rm{3c}}}}$ scaling (Sec.~\ref{subsec:comp_mass}).
These particular time slices are selected to reduce the influence of systematic and statistical errors:
For $t < 16$, the $t$-dependence of the potential from the inelestic states becomes significant, while for $t> 18$, statistical uncertainties grow substantially primarily due to light quarks.
Fig.~\ref{fig:potential} shows potentials extracted at different time slices used to estimate systematic errors associated with inelastic channels and the derivative expansion of the non-local potential.

Fig.~\ref{fig:potential} indicates that the $N$-$\Omega_{\rm 3c}$ potentials in both spin channels are attractive at all distances, exhibiting behavior similar to that observed in the $N$-$\Omega_{\rm{3s}}$ potential \cite{HALQCD:2018qyu}, the $N$-$\phi$ potential \cite{Lyu:2022imf}, and the $N$-${\rm c}\bar{\rm c}$ potentials \cite{Lyu:2024ttm}.
A common feature of these systems is the absence of Pauli exclusion effects between the two hadrons, as they do not share any valence quarks. The attractive $N$-$\Omega_{\mathrm{3c}}$ potentials also share a characteristic two-component structure with $N$-${\rm c}\bar{\rm c}$ potentials \cite{Lyu:2024ttm}, consisting of an attractive core at short distances and an attractive tail at long distances.

We further decompose the potential $V^J_{\rm LO}$ with $J=s$ and $L=0$ into the spin-independent central potential $V_0$ and the spin-dependent one $V_s$ with the $^3{\rm S}_1$ and $^5{\rm S}_2$ channels based on Eq.~(\ref{eq:decompose_spin}),
\begin{equation}
\label{eq:decompose_spin}
    \begin{aligned}
        \vec{s}_{_N}\cdot\vec{s}_{_{\Omega_{\rm 3c}}}&=\frac{1}{2}\left(J(J+1)-\frac{3}{4}-\frac{15}{4}\right),\\
        V_0&=\frac{1}{8}\left(5V_{\rm LO}^{J=2}+3V_{\rm LO}^{J=1}\right),\\
        V_s&=\frac{1}{2}\left(V_{\rm LO}^{J=2}-V_{\rm LO}^{J=1}\right).
    \end{aligned}
\end{equation}

\begin{figure}[t]
    \centering
    \includegraphics[width=0.99\linewidth]{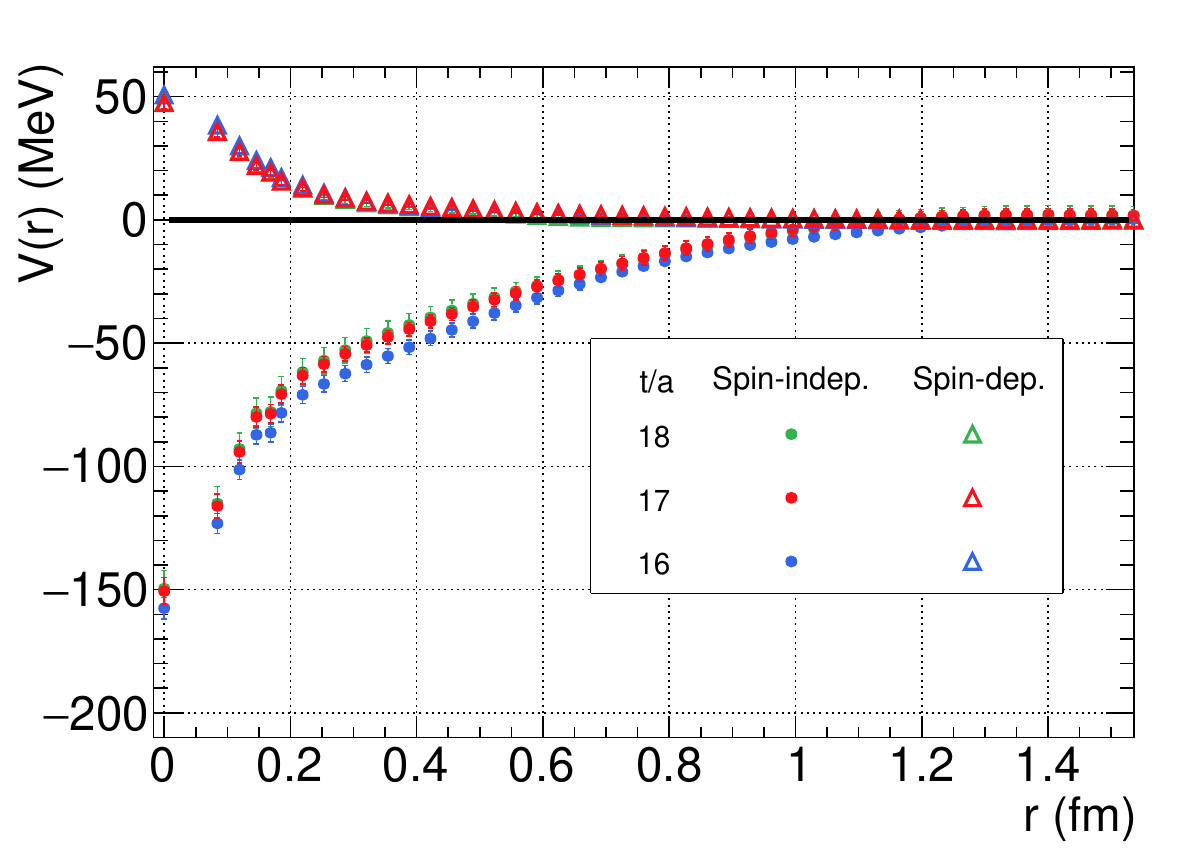}
    \caption{The spin-independent potential $V_0$ (full circle) and spin-dependent potential $V_s$ (open triangle) of $N$-$\Omega_{3c}$, decomposed from the $^3{\rm S}_1$ and $^5{\rm S}_2$ channels, are shown 
    for \colorSchemeChosen.
    }
    \label{fig:pot_spin-dep}
\end{figure}

Fig.~\ref{fig:pot_spin-dep} shows $V_0$ and $V_s$ for \colorSchemeChosen.
 The spin-independent potential provides a significant attractive contribution to the S-wave $N$-$\Omega_{\rm 3c}$ potentials, while the spin-dependent potential acts only at short distances.
Also, {positive values of $V_s$ indicates} $J=s=1$ potential is more attractive than the $J=s=2$ case
because    
  $\left( \vec{s}_{_N}\cdot\vec{s}_{_{\Omega_{\rm 3c}}} \right)_{J=1}  =-5/4$ and $\left( \vec{s}_{_N}\cdot\vec{s}_{_{\Omega_{\rm 3c}}} \right)_{J=2} = 3/4$.   
 
\subsection{Phase shifts and scattering parameters}
\label{sec:phaseshift}
To extract physical observables in infinite spatial volume, we fit the potential using a phenomenological two-range Gaussian form, $V_{\rm fit}(r)=\sum_{i=1,2}A_i\exp\left(-(r/B_i)^2\right)$, which reflects the two-component structure of the $N$-$\Omega_{\rm 3c}$ potentials. 
As an example, 
the uncorrelated fits\footnote{In principle, the data are correlated. A correlated fit requires the number of measurements to be significantly larger than the number of data points~\cite{Dowdall:2019bea}. In the present analysis, however, the total number of configurations is 1600, and the binning procedure (Sec.~\ref{sec:latticeset}) reduces the number of effective samples to 20, while the number of data points is 88. Therefore, a correlated fit is not recommended. Instead, an uncorrelated fit is employed, which provides a conservative estimate of the fitting uncertainties~\cite{Michael:1993yj}.
}
for both channels at $t/a = 17$ over the range $0 < r < 3$~fm are shown in Fig.~\ref{fig:potential}.
The corresponding fitting parameters are listed in Table~\ref{tab:fitparameter}.

\begin{table}[t]
    \centering
\caption{Fitting parameters of the $N$-$\Omega_{\rm 3c}$ potentials 
 for different time slices.}
\label{tab:fitparameter}
\begin{tabular}{crrrr}
    \multicolumn{5}{l}{$^3{\rm S}_1$ channel}\\
    \toprule
    $t/a$ & $A_1$~[MeV] & $A_2$~[MeV] & $B_1$~[fm] & $B_2$~[fm]  \\
    \midrule
        16 & $-118.9(1.6)$& $-85.7(2.6)$ & $0.142(7)$ & $0.633(33)$ \\
        17 & $-118.0(3.0)$ & $-80.0(3.7)$ & $0.135(8)$ & $0.601(37)$ \\
        18 & $-119.7(4.6)$ & $-75.0(8.5)$ & $0.141(12)$ & $0.608(56)$ \\
    \bottomrule
    \multicolumn{5}{l}{}\\
    \multicolumn{5}{l}{$^5{\rm S}_2$ channel}\\
    \toprule
    $t/a$ & $A_1$~[MeV] & $A_2$~[MeV] & $B_1$~[fm] & $B_2$~[fm]  \\
    \midrule
        16 & $-50.5(3.6)$ & $-66.5(3.0)$ & $0.110(15)$ & $0.665(6)$ \\
        17 & $-52.6(2.5)$ & $-60.4(2.5)$ & $0.110(12)$ & $0.612(50)$ \\
        18 & $-53.0(5.2)$ & $-57.8(5.2)$ & $0.113(21)$ & $0.636(67)$ \\
    \bottomrule
\end{tabular}
\end{table}

\begin{figure}[h]
\begin{subfigure}{0.49\textwidth} 
    \centering
        \includegraphics[width=0.99\linewidth]{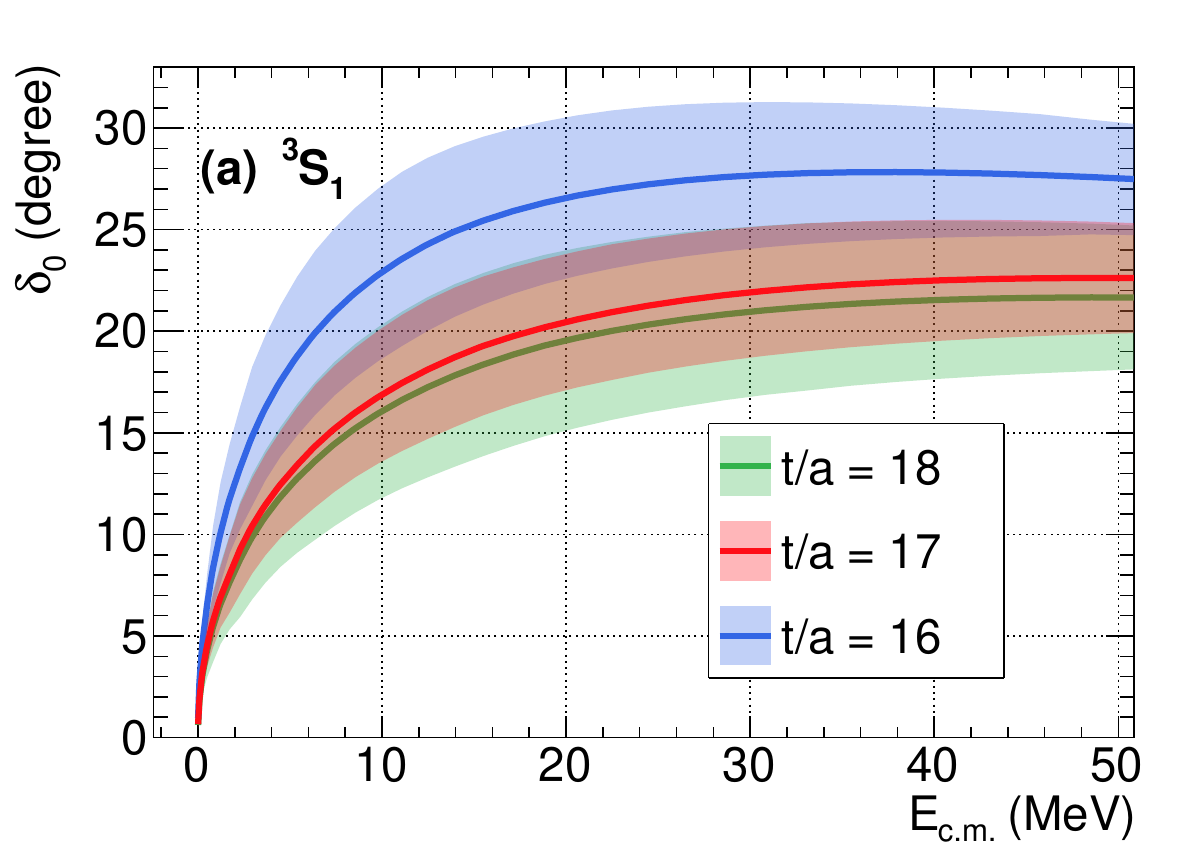}
    \end{subfigure}
    \begin{subfigure}{0.49\textwidth} 
        \centering
            \includegraphics[width=0.99\linewidth]{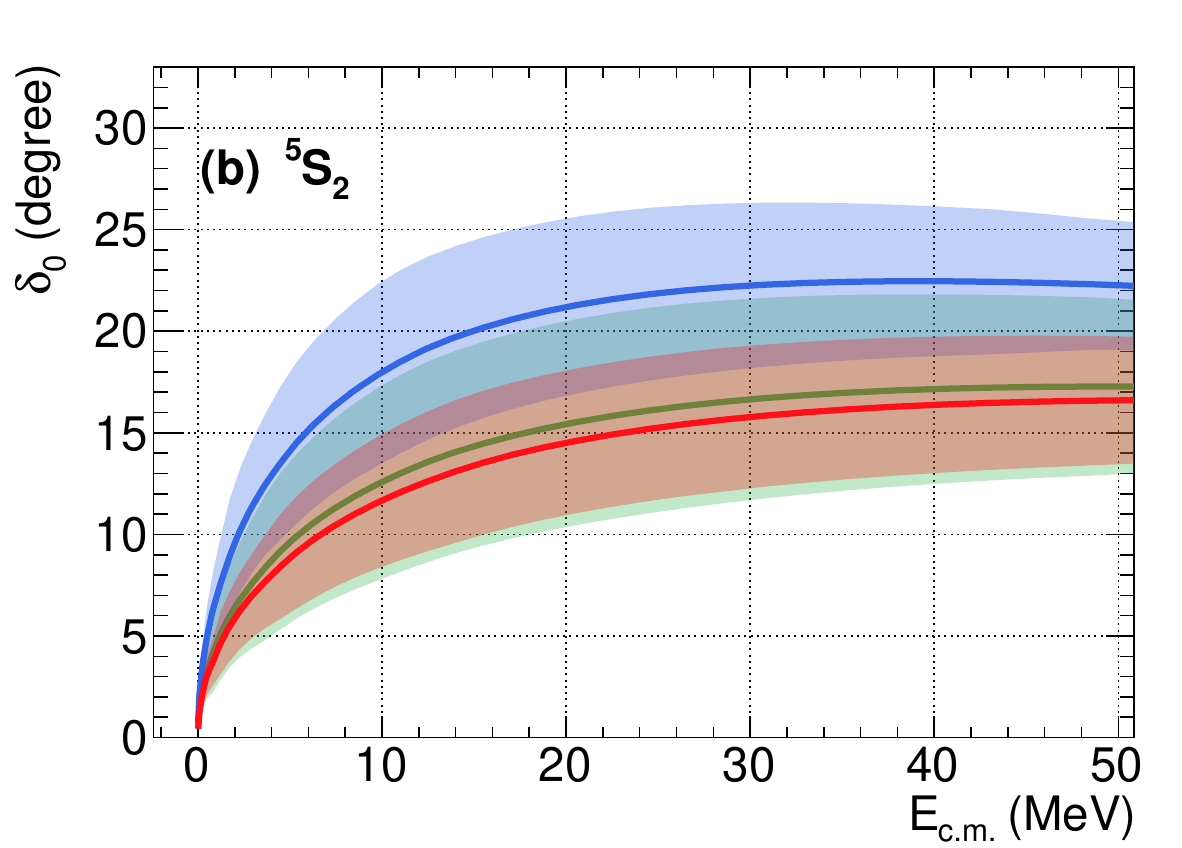}
    \end{subfigure}
    \caption{S-wave $N$-$\Omega_{\rm 3c}$ scattering phase shifts with central values (solid lines) and statistical errors (bands) for 
    \colorSchemeChosen. The panel (a)  for  the $^3{\rm S}_1$ channel and the panel (b) for the $^5{\rm S}_2$ channel.
    }
    \label{fig:phaseshift}
\end{figure}

Fig.~\ref{fig:phaseshift} shows the S-wave $N$-$\Omega_{\rm 3c}$ scattering phase shifts $\delta_0$, calculated as a function of the center-of-mass kinetic energy $E_{\rm{c.m.}}=k^2/2\mu$ for $t/a=16,\ 17 \text{ and }18$. The solid lines represent central values with statistical errors shown as bands, obtained by solving the infinite-volume Schr\"odinger equation with $V_{\mathrm{fit}}(r)$. 
The phase shifts approaching $0^\circ$ in the $k \to 0$ limit corresponds to the absence of bound states below the $N$-$\Omega_{\rm 3c}$ threshold  in either channel, {in accordance with Levinson’s theorem}. 
This contrasts with the quasi-bound state observed in the $^5{\rm S}_2$ $N$-$\Omega_{\rm{3s}}$ channel. The difference arises from the weaker $N$-$\Omega_{\rm 3c}$ potentials compared to the $N$-$\Omega_{\rm{sss}}$ potentials, as will be discussed in more detail in Sec.~\ref{subsec:comp_NO}.

The low-energy scattering parameters are extracted from the effective range expansion (ERE) of the phase shifts, up to the next-{to-}leading-order (NLO),
\begin{equation}
    k\cot\delta_0 = \frac{1}{a_0}+\frac{1}{2}r_{\text{eff}}k^2+\mathcal{O}(k^4),
\end{equation}
where $a_0$ is the scattering length and $r_{\text{eff}}$ is the effective range. Table~\ref{tab:scatteringpara} lists the values obtained from the phase shifts, with mean values calculated at $t/a=17$. The statistical errors are estimated using the jackknife method, while the systematic errors are derived from the results at $t/a=$ 16 and 18. {The effect of higher order terms in the ERE is found to be insignificant for low energy region ($E_{\rm c.m.} \lesssim 50$ MeV).}

\begin{table}
    \centering
\caption{The $N$-$\Omega_{\rm 3c}$ scattering length $a_0$ and effective range $r_{\text{eff}}$ with statistical errors and systematic errors.}
\label{tab:scatteringpara}
    \begin{tabular}{ccc} 
        \toprule
         channel&  $a_0$~[fm]& $r_{\text{eff}}$~[fm]\\ 
        \midrule
            $^3{\rm S}_1$&  $0.56(0.13)\left(^{+0.26}_{-0.03}\right)$& $1.60(0.05)\left(^{+0.04}_{-0.12}\right)$\\ 
            $^5{\rm S}_2$&  $0.38(0.12)\left(^{+0.25}_{-0.00}\right)$& $2.04(0.10)\left(^{+0.03}_{-0.22}\right)$\\
        \bottomrule
    \end{tabular}

\end{table}

\section{Further discussions}
\label{sec:disc}

In this section, we compare the results obtained for the $N$-$\Omega_{\rm 3c}$ system in the present study with those of the analogous systems  $N$-$J/\psi$ and $N$-$\Omega_{\rm 3s}$, which have been investigated previously. 
It is important to note that the $N$-$\Omega_{\rm 3c}$ results are based on the F-conf (with $m_\pi \simeq 137$ MeV), whereas the results for $N$-$J/\psi$ and $N$-$\Omega_{\rm 3s}$  come from the K-conf (with $m_\pi \simeq 146$ MeV). 
Therefore, the comparison is necessarily semi-quantitative. Comparative analyses restricted to the F-conf will be detailed in a forthcoming report.

\subsection{Comparison with \texorpdfstring{$N$-$J/\psi$}{N-Jpsi} potential}
\label{subsec:compNJpsi}

Ref.~\cite{Wu:2024xwy} investigated soft-gluon exchange as the dominant mechanism of the scattering between  $N$ and $J/\psi$, demonstrating its direct connection to the chromo-polarizability of the $J/\psi$. 
Given the observed similarity between the $N$-$\Omega_{\rm 3c}$ and $N$-$J/\psi$ potentials, it is plausible that a similar mechanism also governs the $N$-$\Omega_{\mathrm{3c}}$ system. 

\begin{figure}[t] 
    \centering
    \includegraphics[width=0.49\textwidth]{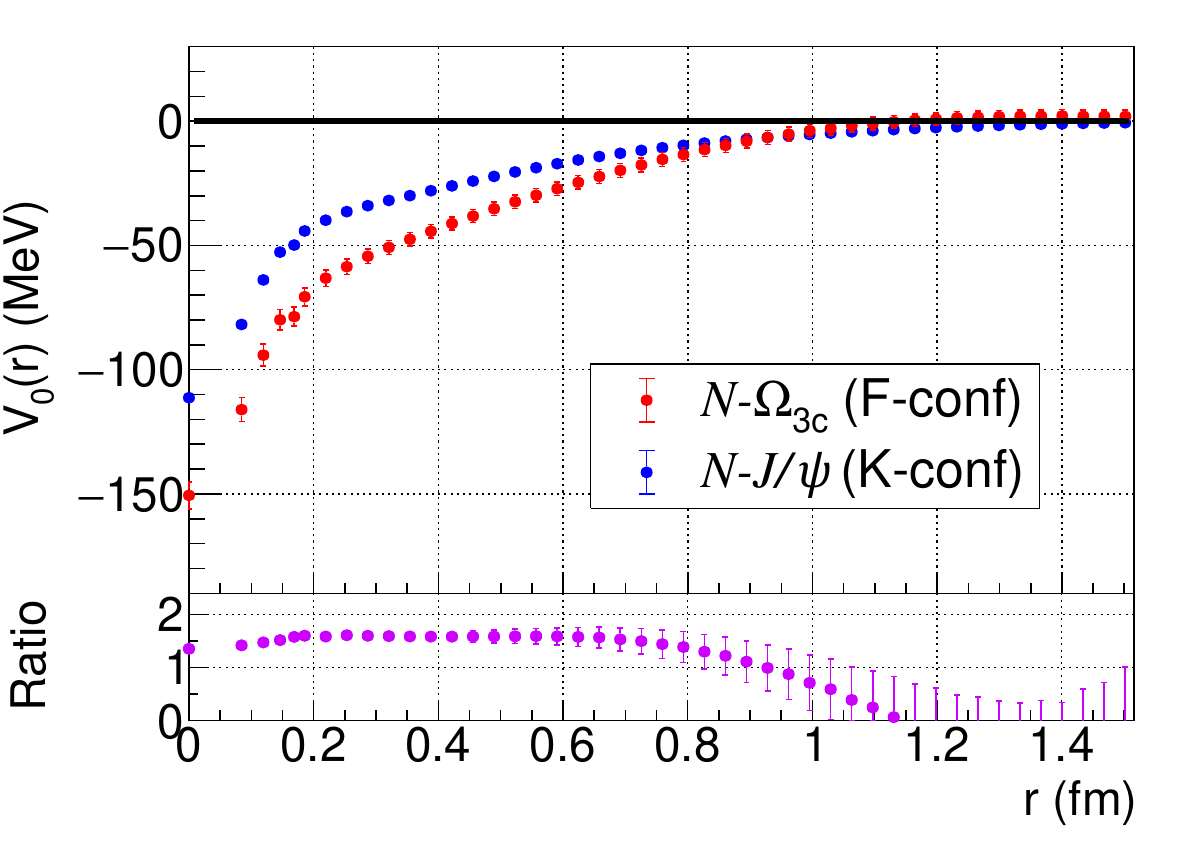}
    \caption{Comparison of the spin-independent potentials for $N$-$\Omega_{\rm 3c}$ from F-conf (at $t/a=17$)
    and $N$-$J/\psi$ from K-conf (at $t/a=13$)~\cite{Lyu:2024ttm}, with the lower panels showing their ratio. 
    }
        \label{fig:comp_V0_Jpsi}
\end{figure}

According to the gluonic van der Waals effective field theory (gWEFT) \cite{Luke:1992tm,Brambilla:2015rqa,Dong:2022rwr}, the coupling of heavy hadrons to external chromo-electric gluon fields is predominantly spin-independent. 
We therefore compare its spin-independent S-wave potential, $V_0$, with that of the $N$-$J/\psi$ system, defined as
$V_{NJ/\psi,0} = {(2V_{NJ/\psi}^{J=3/2} + V_{NJ/\psi}^{J=1/2})}/{3}$,
which is obtained in a manner analogous to Eq.~(\ref{eq:decompose_spin}). The resulting comparison is shown in Fig.~\ref{fig:comp_V0_Jpsi}.

To facilitate an intuitive comparison of interaction strengths, we also plot the ratio of the spin-independent potentials for the $N$-$\Omega_{\rm 3c}$ and $N$-$J/\psi$ systems. As shown in Fig.~\ref{fig:comp_V0_Jpsi}. This ratio exhibits a plateau between $0.2$ and $0.8$~fm, where it remains approximately constant at a value of 1.6.
This suggests that the gluon-mediated dynamics in the $N$-$J/\psi$ and $N$-$\Omega_{\rm 3c}$ systems are qualitatively similar, with the primary difference arising from the magnitude of the chromo-polarizabilities $\beta_{J/\psi}$ and $\beta_{\Omega_{\rm 3c}}$: 
A theoretical estimate gives a ratio 
 $\left(\beta_{\Omega_{\rm 3c}}/\beta_{J/\psi}\right)_{\rm quark\ model} \sim 2.6$ with a 100\% uncertainty \cite{Dong:2022rwr}, 
which does not contradict lattice QCD results, 
 $\left( \beta_{\Omega_{\rm 3c}}/\beta_{J/\psi} \right)_{\mathrm{LQCD}} \simeq 1.6$.

At longer distances, the gluonic interaction effectively reduces to a two-pion exchange process \cite{Bhanot:1979vb,Fujii:1999xn,TarrusCastella:2018php,Wu:2024xwy,Hatsuda:2025djd}.
We have extracted the spatial effective energy~\cite{Lyu:2022imf} to investigate the long-distance behavior of the $N$-$\Omega_{\rm 3c}$ potential.
However, the current  statistical uncertainties prevent us from drawing definitive conclusions about the two-pion dynamics.

\subsection{Comparison with \texorpdfstring{$N$-$\Omega_{\rm{3s}}$}{N-Omegasss} potential}
\label{subsec:comp_NO}
\begin{figure}[t]
   \centering
       \includegraphics[width=0.49\textwidth]{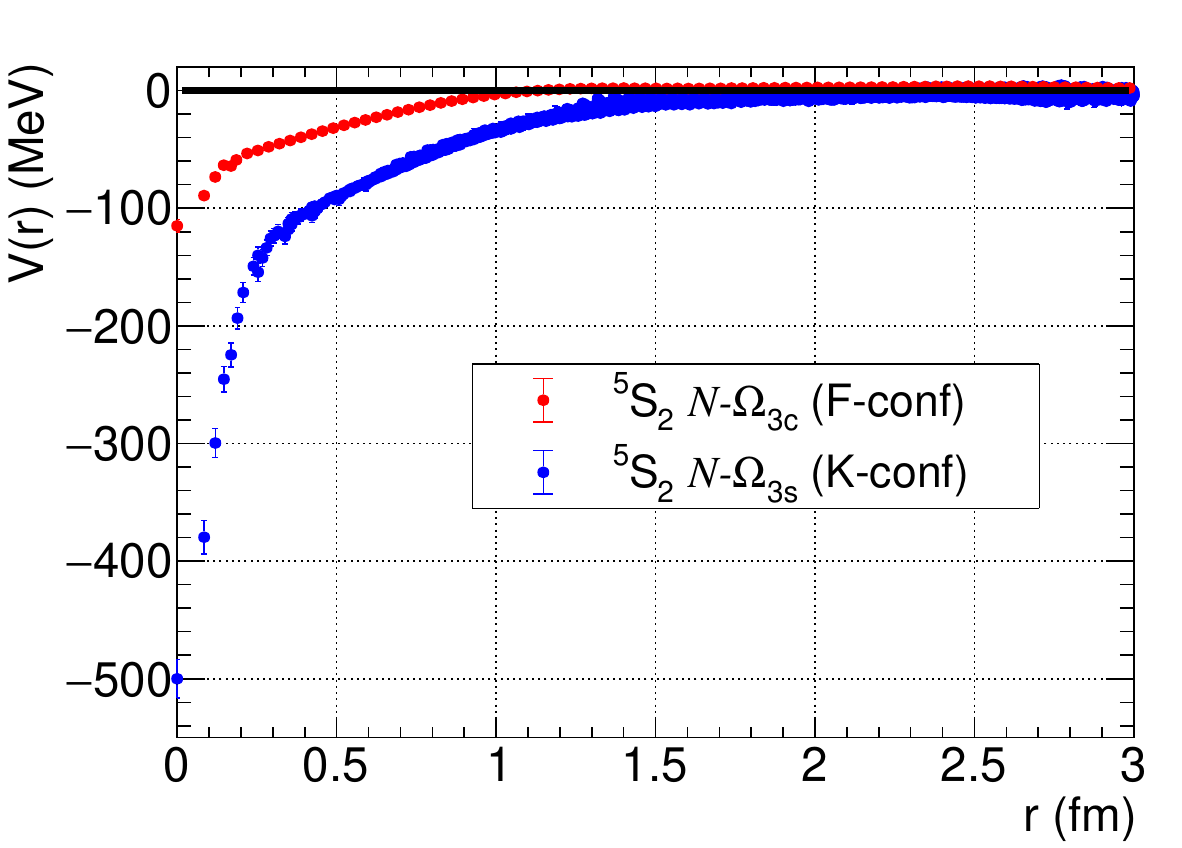}
    \caption{Comparison between potentials for $^5{\rm S}_2\ N$-$\Omega_{\rm 3c}$ from F-conf (at $t/a=17$) and $^5{\rm S}_2\ N$-$\Omega_{\rm 3s}$ from K-conf (at $t/a=14$)~\cite{HALQCD:2018qyu}, showing reduced attraction in $N$-$\Omega_{\rm 3c}$ at short distances, consistent with the phenomenological quark model predictions \cite{Oka:1986fr}. 
    }
    \label{fig:comp_with_NO}
\end{figure}

The $^5{\rm S}_2$ $N$-$\Omega_{\rm{3c}}$ potential obtained in this study using  F-conf exhibits a similar shape but is 2–5 times less attractive compared to the $^5{\rm S}_2$ $N$-$\Omega_{\rm{3s}}$ potential derived with the K-conf \cite{HALQCD:2018qyu}, as shown in Fig.~\ref{fig:comp_with_NO}. Such reduction in attraction is consistent with previous findings from comparisons between the $\Omega_{\rm{3s}}$-$\Omega_{\rm{3s}}$ and $\Omega_{\rm{3c}}$-$\Omega_{\rm{3c}}$ potentials \cite{Lyu:2021qsh}. Several possible origins of this reduction can be considered. 
One possibility is the nature of meson exchange: the exchange of two kaons ($K$) versus two $D$ mesons. 
The former is deeper and longer-ranged due to the lighter mass of the $K$ meson compared to the $D$ meson.
At short distances, the chromo-magnetic interaction, which is inversely proportional to the constituent quark mass ($m^*$) \cite{Oka:1986fr}, may also contribute more significantly to the attraction in the $N$-$\Omega_{\rm{3s}}$ system due to $m_s^* < m_c^*$.
Additionally, a difference in chromo-polarizability, potentially with the ordering $\beta_{\Omega_{\rm 3s}} > \beta_{\Omega_{\rm 3c}}$, could also contribute to the observed behavior.
Further investigation will be required to identify and quantify the source of this difference more precisely.

\subsection{Comparison between different charm masses}
\label{subsec:comp_mass}

\begin{figure}
    \centering
    \includegraphics[width=0.49\textwidth]{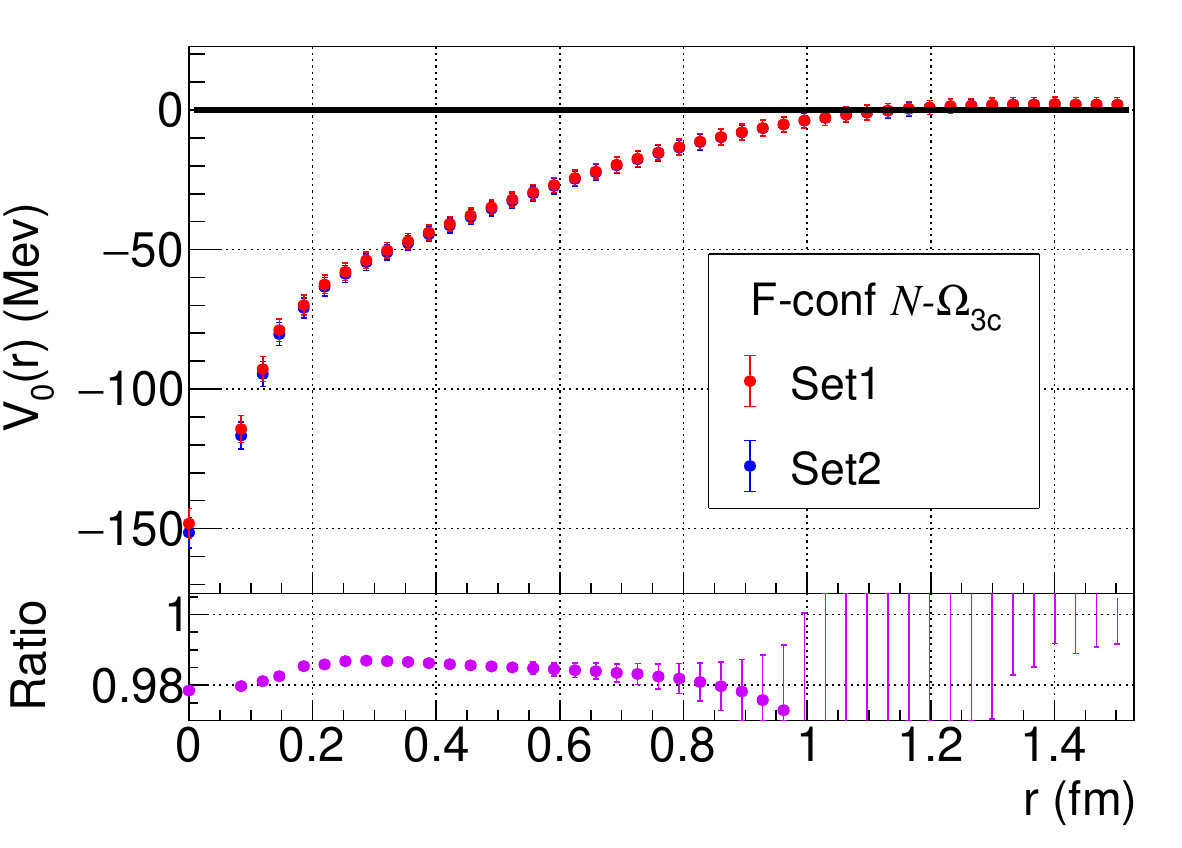}
    \caption{Comparison of $N$-$\Omega_{\rm 3c}$ spin-independent potentials between two RHQ parameter sets with different charm quark masses from F-conf (at $t/a=17$). In the lower figure, Set1 (larger charm quark mass) exhibits reduced potential strength relative to Set2, yielding a ratio below unity with jackknife-estimated errors.
    } 
    \label{fig:comp_sets_V0}
\end{figure}

To investigate charm quark mass dependence of the spin-independent $N$-$\Omega_{\rm 3c}$ potential, we compute the ratio of the potentials using two RHQ parameter sets shown in Table~\ref{tab:intwepolateTwoSet}. Note that identical lattice gauge configurations (F-conf) were used for Set1 and Set2. Although  the two potentials are overlapping with each other for all distances as shown in the upper panel of Fig.~\ref{fig:comp_sets_V0}, their ratio with jackknife-estimated errors shown in the lower panel of Fig.~\ref{fig:comp_sets_V0} exhibits systematic deviation from unity,  indicating the magnitude of the spin-independent $N$-$\Omega_{\rm 3c}$ potential is weaker for larger charm quark mass. 
The obtained ratio between these two sets of potential may imply that
the charm quark mass dependence is described by an approximate $1/m_{_{\Omega_{\rm{3c}}}}$ scaling.

\section{Summary}
\label{sec:summary}

In summary, we investigated the $N$-$\Omega_{\rm 3c}$ system in both $^3{\rm S}_1$ and $^5{\rm S}_2$ channels to explore the interaction between a nucleon and a heavy baryon using $(2+1)$-flavor lattice QCD simulations with physical light quark masses ($m_{\pi} \simeq 137.1\,\text{MeV}$) and a physical charm quark mass defined by
$\left(m_{\eta_c} + 3m_{J/\psi}\right)/4 \simeq 3068.5\,\text{MeV}$.
The masses of the $\Lambda_{\rm c}$-$\Xi_{\rm cc}$ and $\Sigma_{\rm c}$-$\Xi_{\rm cc}$ baryons lie above the $N$-$\Omega_{\rm 3c}$ threshold, allowing for a reliable extraction of the low-energy $N$-$\Omega_{\rm 3c}$ interaction
 in both spin-1 and spin-2 channels via the time-dependent HAL QCD method.
We find that the interaction is attractive in the $^3{\rm S}_1$ ($^5{\rm S}_2$) channel, with a scattering length 
$ a_0 \simeq 0.6\,(0.4)\,\text{fm}$ and $r_{\text{eff}} \simeq 1.6\,(2.0)\,\text{fm}$.
No bound state is observed in the $N$-$\Omega_{\rm 3c}$ system in our lattice QCD results, in contrast to some predictions from quark models~\cite{Huang:2019esu}.

The spin-independent part of the potential, $V_0$, is attractive at all distances and exhibits a two-component structure: a short-range attractive core and a mid-to-long-range attractive tail. On the other hand, the spin-dependent potential $V_s$ is active only at short distances below 0.4 fm. A comparison of the spin-independent potential $V_0$ between the $N$-$\Omega_{\rm 3c}$ and $N$-$J/\psi$ systems at intermediate distances suggests that the difference may be attributed to the different values of the chromo-polarizability between $\Omega_{\rm 3c}$ and $J/\psi$. The $N$-$\Omega_{\rm 3c}$ potential has a similar shape with  the $N$-$\Omega_{\rm{3s}}$ potential with smaller attraction by a factor of 2-5,
which is related to the fact that charm quark mass is much heavier than the strange quark mass. Note that these comparisons are based on lattice data obtained with slightly different light quark masses. A more detailed analysis using the same configurations (only the F-conf) will be presented elsewhere.
We also studied the charm quark mass dependence of the spin-independent part of the $N$-$\Omega_{\rm 3c}$ potential, $V_0$,
and found that the potential is weaker for a larger charm quark mass with approximate  $1/m_{\Omega_{\rm 3c}}$ scaling.

Our lattice QCD study with physical quark masses advances the understanding of heavy-baryon interactions and provides a robust foundation not only for phenomenological investigations of hadron interactions involving charm quarks, but also for experimental studies in high-energy proton-proton and heavy-ion collisions\cite{Chen:2024eaq,Chen:2024aom}.

\vspace{1cm}
We thank members of the HAL QCD Collaboration for stimulating discussions.
The lattice QCD measurements have been performed on Fugaku and HOKUSAI supercomputers at RIKEN.
This work was partially supported by HPCI System Research Project ( hp230075, hp230207, hp240157 and hp240213), the JSPS (Grant Nos. JP18H05236, JP22H00129, JP19K03879, JP21K03555, and JP23H05439), RIKEN Incentive Research Project, ``Program for Promoting Researches on the Supercomputer Fugaku'' (Simulation for basic science: from fundamental laws of particles to creation of nuclei) and (Simulation for basic science: approaching the new quantum era) (Grants No. JPMXP1020200105, JPMXP1020230411), and Joint Institute for Computational Fundamental Science (JICFuS). 
LZ was supported by China Scholarship Council No. 202204910417. And LZ gratefully acknowledges the hospitality and academic support from RIKEN and iTHEMS RIKEN.
YL, TD and TH were supported by Japan Science and Technology Agency (JST) as part of Adopting Sustainable Partnerships for Innovative Research Ecosystem (ASPIRE), Grant No. JPMJAP2318.
This work was also supported  in part by the National Natural Science Foundation of China under contract Nos. 12147101, the Guangdong Major Project of Basic and Applied Basic Research No. 2020B0301030008, the STCSM under Grant No. 23590780100, and the Natural Science Foundation of Shanghai under Grant No. 23JC1400200.

\bibliographystyle{apsrev4-2}
\bibliography{00}

\end{document}